\documentstyle[prl,aps,multicol,graphicx,tabularx]{revtex}

\begin{document}
\bibliographystyle{apsrev}

\preprint{submitted to Phys. Rev. B}

\title{Phonon Scattering and Internal Friction in Dielectric and Metallic Films at Low Temperatures}
\author{P. D. Vu$^1$, Xiao Liu$^2$, and R. O. Pohl$^{3,*}$}
\address{$^1$Soka University of America, Aliso Viejo, CA  92656}
\address{$^2$SFA, Inc., Largo, Maryland 20774}
\address{$^3$LASSP, Cornell University, Clark Hall, Ithaca, New York 14853-2501}
\address{$^{*}$email: pohl@ccmr.cornell.edu}
\date{\today}
\maketitle

\begin{abstract}
We have measured the heat conduction between 0.05~K and 1.0~K of
high purity silicon wafers carrying on their polished faces thin
dielectric films of {\em e}-beam amorphous Si, molecular beam
epitaxial (MBE) Si, {\em e}-beam polycrystalline CaF$_2$, and MBE
CaF$_2$, and polycrystalline thin metallic films of {\em e}-beam
Al, sputtered alloy Al 5056, {\em e}-beam Ti, and {\em e}-beam Cu.
Using a Monte Carlo simulation to analyze the conduction
measurements, we have determined the phonon mean free path within
the films, and found all of them to be much shorter even than in
typical bulk amorphous solids, with no exceptions. We have also
measured the internal friction of these films below 10~K and
found, however, their internal friction at low temperatures
strikingly close to that of amorphous solids, both in magnitude
and in their temperature independence, with the exception of the
MBE Si and alloy Al 5056, whose internal friction is even much
smaller than that of amorphous solids. The internal friction
results indicate the phonon scattering in these thin films is the
same as, or even much less stronger than, in other amorphous
solids, according to the Tunneling Model. Thus, we conclude that
the heat conduction measurements do not support the picture that
the lattice vibrations of these films are glasslike, as had been
surmised earlier for thin metallic films, on the basis of low
temperature internal friction measurements alone [Phys. Rev. B
{\bf 59}, 11767 (1999)].  At the least, the films must contain
additional scattering centers which lead to the very small phonon
mean free path.  Most remarkably, the MBE Si shows the same strong
scattering of thermal phonons as do the other films, while having
the negligible internal friction expected for a perfect film. The
disorder causing the strong scattering of the thermal phonons in
this film is completely unknown. The non-glasslike phonon
scattering phenomena observed here in thin dielectric and
metallic films deserve further investigations.
\end{abstract}

\pacs{PACS numbers: 68.60.-p, 66.70.+f, 63.50.+x, 62.40.+i}

\begin{multicols}{2}

\section{Introduction}

Structure and perfection of thin films on substrates are still
poorly understood. The purpose of the present investigation is to
show that thermal phonons with wavelengths on the order of 100~nm
can be used as very sensitive probes of their disorder. It will
be shown that strong phonon scattering occurs in a large number
of films, including silicon films produced on silicon substrates
by molecular beam epitaxy (MBE). The nature of this disorder is,
however, not understood.

It has recently been shown that the internal friction of
crystalline metal films below 10~K resembles that of amorphous
solids both in magnitude and temperature independence, the
so-called internal friction plateau \cite{33}. A possible
explanation was that crystalline metal films have the same
density of tunneling states as amorphous solids.  In disordered
crystals, such states are called glass-like excitations\cite{94}.
According to the Tunneling Model (TM)\cite{36}, both the low
temperature thermal conductivity below 1~K and the internal
friction plateau below 10~K are determined by the same quantity,
the tunneling strength $C$:
\begin{equation}
C = \frac{\overline P \gamma^2}{\rho v^2_t}, \label{eq:11}
\end{equation}
where $\overline P$ is the uniform spectral density of the
tunneling states, $\gamma$ is their coupling energy to phonons,
$\rho$ is the mass density, and $v_t$ is the transverse sound
velocity. $C$ determines the internal friction plateau through
relaxational scattering of the elastic wave, and the phonon
thermal conductivity through resonant scattering of the thermal
phonons. This quantitative connection between internal friction
and thermal conductivity has been proven in many cases for bulk
amorphous solids and also for disordered crystals, and
constitutes a major proof of the validity of the TM, as reviewed
in Refs. \cite{64} and \cite{76}.

We have recently described a technique by which we can measure
thermal phonon scattering in thin films on substrates \cite{91}.
Using this technique, we have verified the quantitative
connection between internal friction and thermal conductivity
successfully for amorphous silica films\cite{91} and for
crystalline silicon layers which had been disordered by ion
implantation to the point of amorphization\cite{69}. We will use
this technique here for a comparison with the internal friction
on a variety of dielectric and metallic films. If the lattice
vibration of the films are
\end{multicols}
\vbox{
\begin{table}[!h]
\begin{center}
\parbox{140mm}{TABLE I. Preparation parameters of the thin films used
for the heat conduction measurements in this work. The Cu film
has a thin (100~\AA) adhesive layer of Ti between it and the
silicon substrate.  The alloy Al 5056 film contains, by weight,
5.2\% Mg, 0.1\% Mn, and 0.1\% Cr.  The films were deposited onto
either one or both of the wide polished substrate faces (see
Ref.\cite{91}), as indicated in the last column ("one" or "both").
}
\begin{tabularx}{140mm}{>{\small}X>{\small}X>{\small}X>{\small}X>{\small}X>{\small}X>{\small}X}
\hline\hline
  & deposition & base  & substrate  & substrate  & deposition & film \\
 & technique & pressure & temperature & orientation & rate & thickness \\
    &  & (Torr) & ($^\circ$C) &  & (\AA/s) & ($\mu$m) \\ \hline
{\em a}-Si & {\em e}-beam & $2 \times 10^{-7}$ & room & $\langle 100 \rangle$ & 15 & 0.5, one \\
MBE Si & MBE & UHV & 600 & $\langle 100 \rangle$ & 3 & 0.4, one \\
CaF$_2$ & {\em e}-beam & $1 \times 10^{-6}$ & room & $\langle 111 \rangle$ & 10 & 0.1,~both \\
MBE CaF$_2$ & MBE & UHV & 750 & $\langle 111 \rangle$ & 0.14-0.28 & 0.4, one \\
Al & {\em e}-beam & $4 \times 10^{-7}$ & room & $\langle 100 \rangle$ & 15 & 0.2, one \\
 & & & & & & 0.4, one \\
 & & & & & & 0.6, one \\
Al 5056 & sputter & Ar:~$1 \times 10^{-2}$ & room & $\langle 100 \rangle$ & 12 & 0.5, one \\
Ti & {\em e}-beam & $6 \times 10^{-7}$ & room & $\langle 111 \rangle$ & 3 & 0.1,~both \\
Cu & {\em e}-beam & $2 \times 10^{-6}$ & room & $\langle 111 \rangle$ & 6 & 0.1,~both \\
 \hline\hline
\end{tabularx}
\end{center}
\end{table}
}
\begin{multicols}{2}
\noindent indeed glass-like, the phonon scattering should be
determined by the same tunneling strength as determined from the
internal friction.

In addition to the phonon scattering in the metal films studied
previously in internal friction \cite{33}, we will also present
measurements of both internal friction and phonon scattering in
amorphous Si ({\em a}-Si) films, in dielectric crystalline
CaF$_2$ films, and in a crystalline Si film produced by MBE, which
is expected to contain fewer defects than any of the other films.

\section{Experimental Matters}

\subsection{Thin Films}

The thin films for heat conduction measurements were deposited
either on Czochralski-grown, $\langle 111 \rangle$ oriented
silicon substrate surfaces, or on float-zone refined, $\langle 100
\rangle$ oriented ones. For internal friction measurements, the
thin films were deposited onto the double-paddle oscillators,
which were float-zone refined, $\langle 100 \rangle$ oriented. All
substrates were of high purity, and were double-side polished.
Film thickness was determined by calibrated vibrations of a 6~MHz
plano-convex quartz crystal. When possible, it was double-checked
with a step surface profiler.  Details on the films for heat
conduction measurements are contained in Table I.

In order to eliminate surface contaminations to thermal conduction
measurements, samples were, when appropriate, either put through
an RCA clean\cite{61} or cleaned in a hot sulfuric acid solution
\cite{17}. For the internal friction measurements, all films were
deposited directly onto double-paddle oscillators after cleaning
of the substrate by diluted HF solution, except for the MBE Si
film. Because of the stringent cleaning requirements prior to the
MBE deposition in the UHV chamber, which were not suitable for
double-paddle oscillators, the MBE Si film was deposited onto a
wafer from which an oscillator was subsequently fabricated. The
fabrication process involves heating the wafer to 850$^\circ$C for
20~minutes. Thus, the MBE Si film for the internal friction
measurement was considered as annealed. All the other annealing
processes were done in the MOS area of the Cornell
Nanofabrication Facility and were preceded by a stringent RCA
cleaning, as described in Ref.\cite{95}, in order to avoid any
contamination of the silicon which is known to occur during
annealing under regular clean laboratory conditions.  This
annealing process will be referred to in the following as
``MOS-cleaned-and-annealed.''

Since we are primarily interested in the thermal phonon mean free
path in the film from the heat conduction measurement, we would
like to minimize phonon scattering at the film-substrate
interface and at the free surface of the film.  As the
film-substrate interface is expected to be much smoother than the
free surface of the film, we first consider the free surface
roughness of the films studied in this work.  Table II presents
the root-mean-square (RMS) roughness of the free surfaces of the
films studied as determined by atomic force microscopy (AFM).
Table II also presents the dominant thermal phonon wavelength at
1~K based on the Debye speed of sound of the materials
 listed.
The wave length of the thermal phonons is the relevant length. As
can be seen, the RMS roughness is small compared to the length
scale of the thermal phonons; and so, the free surfaces of the
\vbox{
\begin{table}[!h]
\begin{center}
\parbox{80mm}{TABLE II.
Comparison of the free surface RMS roughness as determined by AFM
measurements with the dominant thermal phonon wavelength
$\lambda_{\rm dom}$ at 1~K based on the Debye speed of sound,
$v_D$, for the thin film samples studied.}
\begin{tabularx}{86mm}{>{\small}X>{\small}X>{\small}X>{\small}X>{\small}X} \hline\hline
 & film & ${v_D}^{\rm a}$ & $\lambda_{\rm dom}$ & RMS \\
  & thickness & & at 1~K & Roughness \\
 & ($\mu$m) & (10$^5$~cm/s) & (\AA) & (\AA) \\
\hline
{\em a}-Si & 0.5 & 4.62 & 510 & 5 \\
MBE Si & 0.4 & 5.93 & 650 & $<2$ \\
CaF$_2$ & 0.1 & 4.10 & 450 & 25 \\
MBE CaF$_2$ & 0.4& 4.10 & 450 & 7 \\
Al & 0.2 & 3.42 &370 & 40 \\
 & 0.4 & 3.42 & 370 & 20 \\
 & 0.6 & 3.42 & 370 & 40 \\
Al 5056 & 0.5 & 3.42 & 370 & 40 \\
Ti & 0.1 & 3.48 & 380 & 10 \\
Cu & 0.1 & 2.78 & 300 & 10 \\
 \hline\hline
\end{tabularx}
\end{center}
$^{\rm a}$ Taken from Ref.\protect\cite{3} or estimated using
Eq.~\ref{eq:31} with $v_t$ listed in Table III.
\end{table}
} films should have little effect on the overall scattering of
thermal phonons and thus on the overall conclusions of this work.
In section III C, we will show experimental evidence that phonon
scattering at the surfaces/interfaces is indeed negligible
relative to that in the films, by studying films thickness
dependence.

\subsection{Methods}

In contrast to the conventional thermal conductivity measurements
on bulk metals, our thermal method, as applied to metal films,
has the advantage of solely determining the phonon heat transport
in the form of a phonon mean free path in the film.  In our
investigation the metal films act primarily as phonon scatterers
rather than as heat conductors because of their relatively small
thickness to that of the substrate. At low temperatures, most of
the heat in normal metals is carried by electrons.  This thermal
conductivity can be calculated using the Wiedemann-Franz-Lorenz
law (reviewed in Ref.\cite{96}) and an appropriate electrical
resistivity. Because heat transport is parallel to the
film-substrate interface, the amount of heat carried by the film
or the substrate depends on their relative thermal resistance,
which is the inverse of thermal conductivity multiplied by length
and divided by cross-sectional area. Since the film and substrate
have the same length and width and differ only in thickness, a
comparison of the products of thermal conductivity and thickness
is enough to determine which carries most of the heat, which is
shown in Fig.~\ref{fig1} using a 0.2~$\mu$m thick Cu film as an
example. The low temperature thermal conductivity of the Cu film
was determined with the \vbox{
\begin{figure}[!h]
\begin{center}
\includegraphics[scale=0.5]{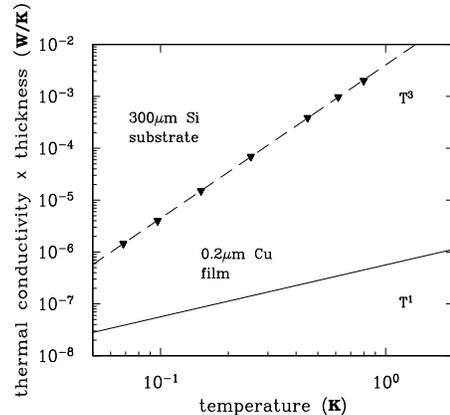}
\caption[a]{Measured thermal conductivity of a high purity Si
substrate (large faces polished and thin faces sandblasted, see
Appendix) multiplied by its thickness, 300~$\mu$m: solid
triangles; and the calculated electronic thermal conductivity of a
Cu film (see text) multiplied by its thickness, 0.2~$\mu$m: solid
line. At 50~mK, only about 5\% of the heat is carried in the film
(by electrons); at higher temperatures, the percentage drops as
$T^{-2}$.} \label{fig1}
\end{center}
\end{figure}
} Wiedemann-Franz-Lorenz law using a room temperature electrical
resistivity of $1.6 \times 10^{-6}$~$\Omega$cm and a residual
resistivity ratio of 2, measured in our laboratory on similar
films \cite{33,cahill-1990}. Fig.~\ref{fig1} demonstrates that
the substrate phonons are the dominant heat carriers, primarily
because the thickness of the substrate is so much greater than
that of the film.  The same is true for the other metal films
studied in this work, even more so when they become
superconducting in the temperature range investigated here.

Thermal phonon mean free paths were determined, using a Monte
Carlo (MC) simulation, from heat conduction measurements between
0.05 and 1.0~K by the technique described in Ref.\cite{91},
denoted as $\ell_{\rm film(HC)}$ in the following. We mention
briefly that additional scattering mechanisms such as scattering
from free surface roughness can be included in the simulations,
should that become necessary.  For those who wish to determine
$\ell_{\rm film(HC)}$ from heat conduction measurements below 1~K
without having to resort to performing their own simulations, we
provide information in the Appendix, using the results of our MC
simulations on a film-substrate sample with dimensions as
typically used in our work.

We can also predict the phonon mean free path from internal
friction measurements, denoted as $\ell_{\rm film(TM)}$, if we
assume that the film has the low energy excitations that are
common in amorphous solids (and no other scattering centers)
within the TM model.  In the present work, the low-temperature
internal friction of thin films is measured with double-paddle
oscillators vibrating in their antisymmetric mode at $\sim
5.5$~kHz, which have exceptionally small background damping as
described previously\cite{35,106}. Thin films increase the
internal friction of the paddle oscillator, $Q_{\rm
paddle}^{-1}$.  From this, the internal friction of the film,
$Q_{\rm film}^{-1}$, is determined by\cite{35}
\begin{equation}
Q_{\rm film}^{-1} = \frac{G_{\rm sub} t_{\rm sub}}{3 G_{\rm film} t_{\rm film}} (Q_{\rm paddle}^{-1} - Q_{\rm sub}^{-1}),
\label{eq:21}
\end{equation}
where $t$ and $G$ are thicknesses and shear moduli of substrate
and film, respectively, and $Q_{\rm sub}^{-1}$ is the internal
friction of the bare paddle (including the mounting losses).
$G_{\rm film}$ is assumed to be equal to that of the bulk
material \cite{33,35}. The specific model used to obtain
$\ell_{\rm film(TM)}$ from the internal friction of a film is the
TM, originally proposed by Anderson, et al.\cite{62}, and
independently by Phillips\cite{63}, expanded for elastic
measurements by J\"{a}ckle\cite{97}. The TM connects the thermal
phonon mean free path, $\ell$, with the internal friction
plateau, $Q_0^{-1}$, as follows.  From Ref.\cite{76}, the
expression for the thermal conductivity $\Lambda$ is
\begin{equation}
\Lambda = \frac{1}{3} C_v v_{D} \ell = \frac{2.66 k_B^3}{6\pi
\hbar^2} \frac{\pi}{2Q_0^{-1} v_t} T^2 \label{eq:27}
\end{equation}
where, in the gas-kinetic picture, $C_v$ is the low temperature
specific heat per unit volume, $v_D$ is the Debye speed of sound,
$k_B$ is Boltzmann's constant, $\hbar$ is Planck's constant, and
$T$ is the temperature.  Note that
\begin{equation}
Q_0^{-1} = \frac{\pi}{2} C, \label{eq:22}
\end{equation}
where $C$ is defined in Eq.~\ref{eq:11}. Substituting for $C_v$
within the Debye model of the phonon spectrum\cite{68},
Eq.~\ref{eq:27} becomes
\begin{equation}
\ell = (1.59 \times 10^{-12} [{\rm s\thinspace K}])
\frac{v_t}{Q_0^{-1}} T^{-1}, \label{eq:33}
\end{equation}
assuming the empirical relation\cite{3,5}:
\begin{equation}
v_t \simeq 0.9 v_D, \label{eq:31}
\end{equation}
where [s\thinspace K] are units of second and Kelvin. These
equations provide the means to predict (within the TM) what
$\ell_{\rm film(TM)}$ should be if the internal friction plateau
of the film, ${Q_0^{-1}}_{\rm film}$, is known. To repeat,
Eq.~\ref{eq:33} assumes that the internal friction plateau is due
to the presence of glassy states in the film, and that no defects
other than the glass-like excitations scatter the thermal
phonons.  The validity of these assumptions will be tested for
the films investigated here by comparing $\ell_{\rm film(HC)}$
with $\ell_{\rm film(TM)}$.

\section{Results and Discussion}

\subsection{Silicon Films}

Since we are searching for tunneling states in thin films, we
start with {\em e}-beam {\em a}-Si, a highly disordered film
known \vbox{
\begin{figure}[!h]
\begin{center}
\includegraphics[scale=0.5]{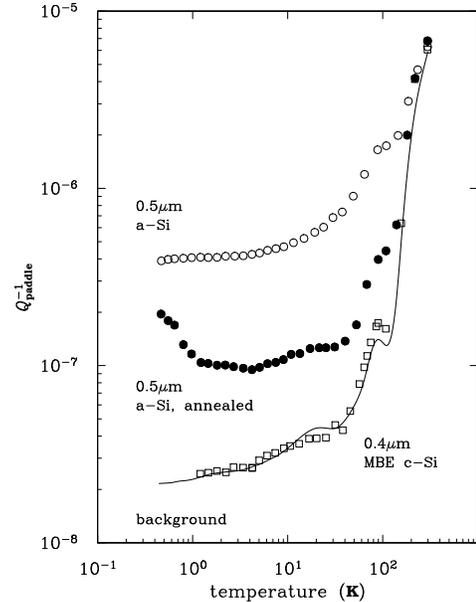}
\caption[a]{The internal friction of a bare double-paddle
oscillator (solid curve ``background'') and of such oscillators
carrying {\em e}-beam {\em a}-Si and MBE Si films. Note the
negligible effect of the MBE film. The annealing of the {\em
e}-beam {\em a}-Si film was done at 700$^\circ$C for 1~hr under
the MOS-cleaned-and-annealed condition (see text). The MBE Si
film had been annealed at 850$^\circ$C for 20~min (see text).}
\label{fig2}
\end{center}
\end{figure}
\begin{figure}[!h]
\begin{center}
\includegraphics[scale=0.5]{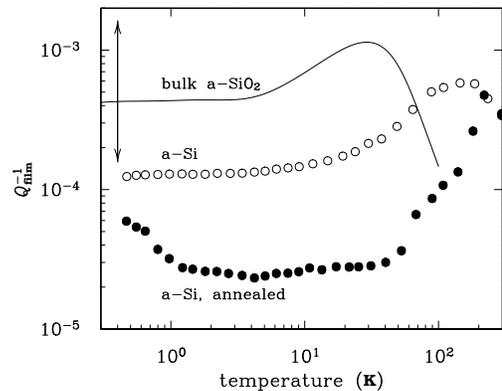}
\caption[a]{Internal friction of an {\em e}-beam {\em a}-Si film,
before and after annealing, compared to that of bulk {\em
a}-SiO$_2$ (solid curve). The bulk {\em a}-SiO$_2$ data, measured
at 4.5~kHz, are taken from J.E. Van Cleve, Ph.D. thesis, Cornell,
published in Ref.~\cite{64}. The double-headed vertical arrow
indicates the range of the temperature-independent internal
friction plateau, measured on a wide range of bulk amorphous
solids as reviewed in Ref.~\cite{76}.} \label{fig3}
\end{center}
\end{figure}
}
to have such states\cite{35}, and will compare it with a
crystalline silicon film produced by MBE which is expected to be
a simple extension of the silicon lattice.  Fig.~\ref{fig2} shows
the internal friction of a bare double paddle oscillator, called
``background,'' and of the same kind of oscillator carrying the
films. As expected, the MBE film has negligible internal
friction, while the {\em e}-beam {\em a}-Si films, both
as-deposited and annealed, lead to a considerable increase of the
internal friction. The MBE Si was measured only in the annealed
state as explained in Section II A.

From the change of the internal friction of the paddle carrying
the films, the internal friction of the films, $Q^{-1}_{\rm
film}$, can be determined using Eq.~\ref{eq:21}, and is compared
to that of bulk {\em a}-SiO$_2$ in Fig.~\ref{fig3}. The annealing
of the {\em e}-beam film causes almost an order of magnitude
reduction in $Q^{-1}_{\rm film}$, while the $Q^{-1}_{\rm film}$
of the annealed MBE film is too small to be determined. Annealing
at 700$^\circ$C for 1~hr leads to almost complete crystallization
of a $0.5$~$\mu$m thick {\em e}-beam {\em a}-Si film \cite{98}.
The internal friction confirmed that $\sim$ 90\% of the low energy
excitations had been removed. Below 1.0~K, however, the internal
friction increased, indicative of a contamination in {\em c}-Si
\cite{95}, although the most stringent MOS-cleaned-and-annealed
process, as described in Section II A, was strictly followed.
Most probably, an impurity was trapped on the substrate surface
during mounting in the {\em e}-beam evaporator which is located
outside the MOS area.  This impurity, trapped underneath the {\em
a}-Si film, survived the MOS cleaning and led to the
contamination during the annealing. In contrast, there is no such
problem for the MBE Si film, and hence no contamination is
observed.

Table III summarizes ${Q_0^{-1}}_{\rm film}$, $v_t$, $\ell_{\rm
film(TM)}(T)$ given by Eq.~\ref{eq:33}, for the films presented
here and below. This en- \vbox{
\begin{table}[!h]
\begin{center}
\parbox{80mm}{TABLE III.
The internal friction plateau ${Q_0^{-1}}_{\rm film}$, the
transverse speed of sound $v_t$, and the thermal phonon mean free
path $\ell_{\rm film(TM)}$($T$) predicted by Eq.~\ref{eq:33},
where $T$ is measured in Kelvin.}
\begin{tabularx}{86mm}{>{\small}X>{\small}X>{\small}X>{\small}X}
\hline\hline
    & ${Q_0^{-1}}_{\rm film}$ & $v_t$ & $\ell_{\rm film(TM)}$($T$) \\
        &  & (10$^5$~cm/s) & $(\mu$m) \\
\hline
{\em a}-Si & $1.3 \times 10^{-4}$ & 4.16$^{\rm b}$ & $50.9/T$ \\
annealed {\em a}-Si & $2.4 \times 10^{-5}$ & 5.33$^{\rm c}$ & $353/T$ \\
MBE Si & negligible & 5.33$^{\rm c}$ & $^{\rm e}$ \\
CaF$_2$ & $6.0 \times 10^{-5}$ & 3.69$^{\rm c}$ & $97.8/T$ \\
Al & $1.0 \times 10^{-4}$\thinspace$^{\rm a}$ & 3.04$^{\rm c}$ & $48.3/T$ \\
Al 5056 & $1.0 \times 10^{-5}$\thinspace$^{\rm a}$ & 3.04$^{\rm c}$ & $483.4/T$ \\
Ti & $2.0 \times 10^{-4}$\thinspace$^{\rm a}$ & 3.13$^{\rm d}$ & $24.9/T$ \\
Cu & $5.3 \times 10^{-4}$\thinspace$^{\rm a}$ & 2.50$^{\rm c}$ & $7.5/T$ \\
\hline\hline
\end{tabularx}
\end{center}
$^{\rm a}$ Taken from Ref. \cite{33}

$^{\rm b}$ Taken from Ref. \cite{35}

$^{\rm c}$ Taken from Ref. \cite{5}

$^{\rm d}$ Taken from Ref. \cite{100}

$^{\rm e}$ There is no ${Q_0^{-1}}_{\rm film}$ value for the MBE
Si film because the internal friction of this film was not
detectable (see Fig.~\ref{fig2}), and hence $\ell_{\rm film(TM)}$
is expected to be very long.

\end{table}
} ables us to compare the internal friction and the phonon mean
free path directly within the TM.

Fig.~\ref{fig4} shows $\ell_{\rm film(HC)}$ in the MBE Si and
{\em e}-beam {\em a}-Si films both in as-deposited and annealed
states obtained from the heat conduction measurements. The dotted
line represents $\ell_{\rm film(TM)}$ of the {\em e}-beam {\em
a}-Si based on the internal friction measurement (see Table III).
For {\em e}-beam {\em a}-Si, $\ell_{\rm film(HC)}$ is
significantly smaller than $\ell_{\rm film(TM)}$ which assumes
that the scattering occurs by tunneling states alone. Evidently,
$\ell_{\rm film(HC)}$ cannot be used to test for the existence of
glassy excitations in {\em e}-beam {\em a}-Si film. In addition
to the tunneling states, other scattering centers must be
present. The situation may be similar to {\em e}-beam {\em
a}-SiO$_2$ films in which the additional thermal phonon
scattering was explained by cracks or voids\cite{91}. It is well
known that {\em a}-Si films produced by {\em e}-beam evaporation
contain similar defects\cite{71,72,73}. After annealing, only a
small increase of $\ell_{\rm film(HC)}$ above 0.3~K and even a
decrease below that temperature can be seen in Fig.~\ref{fig4}.
This change of $\ell_{\rm film(HC)}$ upon annealing, which cannot
be explained within the TM, is interpreted as resulting from a
combination of a decreased scattering by the tenfold smaller
number of tunneling states, an increased scattering by the
contaminants, and possibly a change in scattering from the
defects of unknown nature which had  been noticed already in the
film prior to annealing. Obvi-\vbox{
\begin{figure}[!h]
\begin{center}
\includegraphics[scale=0.5]{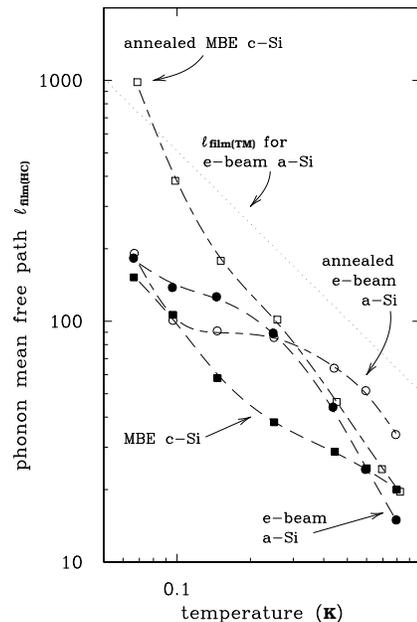}
\caption[a]{Phonon mean free path $\ell_{\rm film(HC)}$ of
as-deposited and annealed silicon films. As-deposited {\em
e}-beam {\em a}-Si: solid circles; annealed (700$^\circ$C, 1~hr)
{\em e}-beam {\em a}-Si: Open circles; as-deposited MBE Si: solid
squares; annealed MBE Si (500$^\circ$C, 1~hr followed by
700$^\circ$C, 1.5~hrs): Open squares. The dotted line is
$\ell_{\rm film(TM)}(T)$, the TM prediction based on the internal
friction plateau of the as-deposited {\em e}-beam {\em a}-Si.
Dashed lines are guides for the eye.} \label{fig4}
\end{center}
\end{figure}
} ously, these results cannot be used to extract any knowledge
about the annealing of these unknown defects.  They do, however,
provide further evidence for the presence of the contaminants
introduced into crystalline silicon during annealing under all
but the most stringent conditions, thus further emphasizing the
need for their identification and control \cite{95}.

Very surprisingly, Fig.~\ref{fig4} shows that the phonon
scattering in MBE Si film both before and after annealing is very
large as well. Since the contamination that plagued the {\em
e}-beam {\em a}-Si films is not an issue here, secondary ion mass
spectroscopy (SIMS) was performed by R. Reedy at the National
Renewable Energy Laboratory (NREL) in order to identify other
possible chemical impurities. The following chemical elements
were detected: boron,  $< 10^{16}$~cm$^{-3}$; nitrogen, $\sim 3
\times 10^{16}$~cm$^{-3}$; carbon, $\sim 3 \times
10^{17}$~cm$^{-3}$; and hydrogen, $\sim 2 \times
10^{18}$~cm$^{-3}$.  These concentrations were similar in the MBE
film and the silicon substrate.  Only the oxygen contents
differed between substrate ($\sim 5 \times 10^{17}$~cm$^{-3}$)
and film ($\sim 5 \times 10^{18}$~cm$^{-3}$).  For all these
detected impurities, an anneal (500$^\circ$C, 1~hr followed by
700$^\circ$C, 1.5~hr) caused no measurable change of their
concentrations. Since no evidence for such scattering was
observed on bare silicon samples from the same batch as  used in
these experiments\cite{91}, the only possible impurity scatterer
is the oxygen. But the tenfold increase of oxygen in the MBE film
should not lead to a noticeable decrease of the experimental mean
free path $\ell$ (defined in the Appendix, Eq.~\ref{eq:2}), and
thus to a decrease of $\ell_{\rm film(HC)}$ given the relatively
small thickness of the film, unless the oxygen in the film
somehow acts as a much stronger scattering center. Phonon
scattering in oxygen-doped silicon has been found to depend on
heat treatment\cite{104,105} at frequencies in excess of $\sim
300$~GHz. But, phonons in this frequency range carry heat
predominantly above 3~K, and no evidence exists for phonon
scattering by oxygen at lower frequencies (corresponds to
$T<1$~K).

In an attempt to detect any evidence for structural disorder in
this MBE film, an x-ray diffraction (XRD) analysis was performed
by M. Sardela and D. Cahill at the University of Illinois
(Champaign-Urbana). High resolution open-detector scans around
the Si(004) peak showed no difference between measurements
conducted on the film side and on the substrate side of the MBE
Si film-substrate sample.  Full width at half maximum values were
found to be almost identical on both sides, and no diffuse
scattering or any disorder feature on the film side was seen. In
addition, a triple axis reciprocal space map around the Si(004)
peak on the film side also could not detect any diffuse
distribution nor asymmetry of the Si peak. These observations
speak against the existence of grains with different orientations
which might cause the phonon scattering.

The near $T^{-1}$ temperature dependence may be suggestive of
scattering by sessile dislocations as reported, for example, by
Wasserb\"{a}ch in plastically deformed bulk copper, niobium, and
tantalum\cite{49}.  If we assume that the coupling between
dislocations and phonons is similar, the dislocation density in
the MBE film would have to range between $10^{10}$ and
$10^{13}$~cm$^{-2}$, which seems rather high for MBE silicon. At
this point, no search for dislocations in this MBE film has been
undertaken.

MOS-cleaning-and-annealing of the MBE film as described above
leads to an increase of $\ell_{\rm film(HC)}$ shown in
Fig.~\ref{fig4}, although it remains below the mean free path
predicted even for {\em e}-beam {\em a}-Si, except at the lowest
temperature. Thus, noticeable disorder, other than the tunneling
states, remains in the MBE Si film even after annealing, although
the internal friction of the annealed MBE Si film shown in
Fig.~\ref{fig2} give no evidence for any low energy excitations.
Thus, the only firm conclusion we can draw at this point is that
the defects in the MBE film are not glass-like.

\subsection{CaF$_2$ Films}

As was just shown, crystalline films produced by crystallizing an
{\em a}-Si film or by MBE deposition show little or no evidence
for tunneling states in low temperature internal friction.  It is
therefore surprising that a 0.6~$\mu$m thick film of crystalline
{\em e}-beam CaF$_2$ increases the damping of the double paddle
oscillator by more than one order of magnitude, see
Fig.~\ref{fig5}. The internal friction of the film itself, as
compared with that of {\em a}-SiO$_2$ shown in Fig.~\ref{fig6}, is
nearly temperature independent and close to the range found for
all amorphous solids studied to date (with the exception of
certain hydrogenated {\em a}-Si films, as discussed in
Ref.\cite{35}). Thus, the large internal friction observed
previously in polycrystalline metal films\cite{33} apparently
also occurs in some crystalline dielectric films. Assuming that
its cause is glass-like excitations, we can again predict an
$\ell_{\rm film(TM)}$ for the CaF$_2$ film, shown as the dotted
line in Fig.~\ref{fig7}. The measured $\ell_{\rm film(HC)}$ for an
identical CaF$_2$ film, also shown in Fig.~\ref{fig7}, is more
than two orders of magnitude smaller than predicted by the TM. We
also measured the phonon mean free path of another crystalline
CaF$_2$ film which was prepared by MBE technique at 750$^\circ$C
substrate temperature to improve its structure. However, the
internal friction of the MBE CaF$_2$ film cannot be measured
because of the difficulty of preparing such a film on silicon
paddle oscillators and meeting the requirements of special
cleaning and fabrication at the same time. The $\ell_{\rm
film(HC)}$ of the MBE CaF$_2$ film, though larger than that of
the {\em e}-beam one, is still smaller than that predicted by the
TM for the {\em e}-beam CaF$_2$ film. Furthermore, it is smaller
than that of an {\em a}-SiO$_2$ film prepared by wet-thermal
oxidation, in which the structure is much improved in comparison
with {\em e}-beam {\em a}-SiO$_2$, and the phonon scattering is
determined solely by the glassy excitations. The \vbox{
\begin{figure}[!h]
\begin{center}
\includegraphics[scale=0.5]{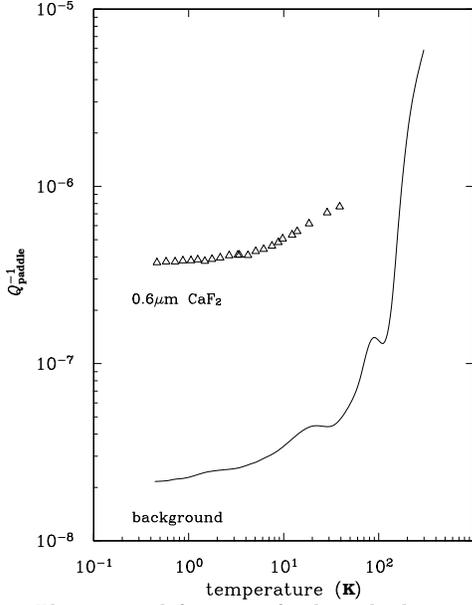}
\caption[a]{The internal friction of a bare high purity silicon
double-paddle oscillator (solid curve ``background'') and of such
a paddle carrying the {\em e}-beam CaF$_2$ film on the polished
silicon surface.} \label{fig5}
\end{center}
\end{figure}

\begin{figure}[!h]
\begin{center}
\includegraphics[scale=0.5]{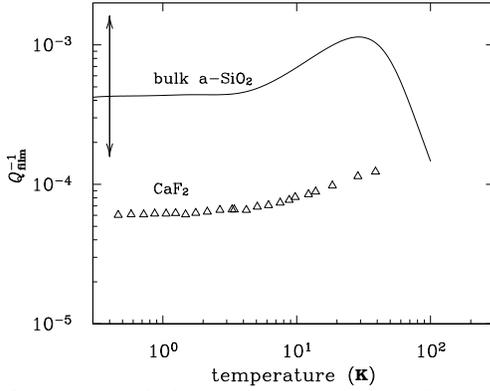}
\caption[a]{Internal friction of the {\em e}-beam CaF$_2$ film
compared to that of bulk {\em a}-SiO$_2$ (solid curve).  The bulk
{\em a}-SiO$_2$ data, measured at 4.5~kHz, is taken from J.E. Van
Cleve, Ph.D. thesis, Cornell, published in Ref.\cite{64}. The
double-headed vertical arrow indicates the range of the
temperature-independent internal friction plateau, measured on a
wide range of bulk amorphous solids as reviewed in Ref.\cite{76}.}
\label{fig6}
\end{center}
\end{figure}
} $\ell_{\rm film(HC)}$ of the thermal {\em a}-SiO$_2$ film
agrees perfectly with that of the TM's prediction, just as one
would expect for bulk {\em a}-SiO$_2$, as shown in
Fig.~\ref{fig7} (see Ref.\cite{91} for details). The $\ell_{\rm
film(HC)}$ of the MBE CaF$_2$ film locates between those of the
thermal {\em a}-SiO$_2$ film and a macroscopically as well as
microscopically disordered {\em e}-beam {\em a}-SiO$_2$ film, in
which the thermal phonon scattering is not dominated by the
tunneling states, also shown in Fig.~\ref{fig7}. We suggest that
this MBE CaF$_2$ film, al- \vbox{
\begin{figure}[!h]
\begin{center}
\includegraphics[scale=0.5]{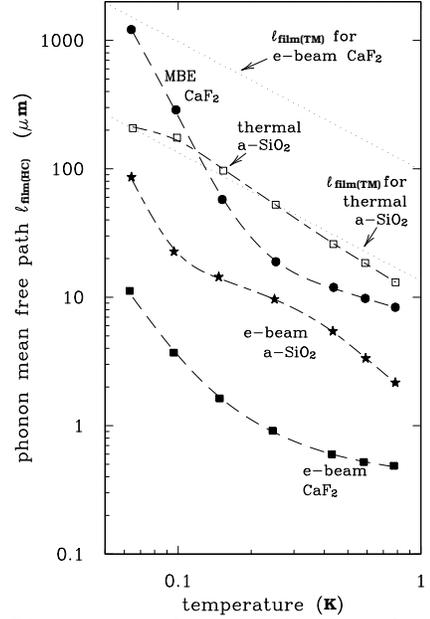}
\caption[a]{Phonon mean free path $\ell_{\rm film(HC)}$ of two
different CaF$_2$ films. MBE CaF$_2$: solid circles; {\em e}-beam
CaF$_2$: solid squares. The dotted lines are the TM prediction
based on internal friction measurements for {\em e}-beam CaF$_2$
and for thermal {\em a}-SiO$_2$, respectively. Data for the
thermal (open squares) and {\em e}-beam (solid stars) {\em
a}-SiO$_2$ films are taken from Ref.\cite{91}. For thermal {\em
a}-SiO$_2$, good agreement is shown between $\ell_{\rm film(HC)}$
and $\ell_{\rm film(TM)}$, which had also been found in
ion-implanted silicon \cite{69}, as mentioned in Section I. Dashed
lines are guides for the eye.} \label{fig7}
\end{center}
\end{figure}
} though probably more highly ordered than an {\em e}-beam one,
contains disorder because the pseudomorphic epitaxial growth is
known to break down at film thickness exceeding
10~nm\cite{101,102}, leading to structural relaxation. Internal
stresses are also expected to result from differential thermal
contraction as the sample is cooled from the deposition
temperature. There is, however, no convincing evidence for
glass-like excitations in this CaF$_2$ film.  As in the {\em
e}-beam {\em a}-Si film, some unknown scattering process masks
the effect of the glass-like excitations, if they exist at all.

\subsection{Metal Films}

Fig.~\ref{fig8} shows $\ell_{\rm film(HC)}$ for three {\em
e}-beam Al films, 0.2, 0.4, and 0.6~$\mu$m thick.  The absence of
any significant dependence on the film thickness validates the
assumption used in our analysis that the scattering occurs
predominantly within the films and not at the interfaces, an
assumption which so far had been based only on the smoothness
observed on the free surfaces as listed in Table II.  The dotted
line is the prediction for $\ell_{\rm film(TM)}$ based on the
internal friction of the {\em e}-beam Al film reported \vbox{
\begin{figure}[!h]
\begin{center}
\includegraphics[scale=0.5]{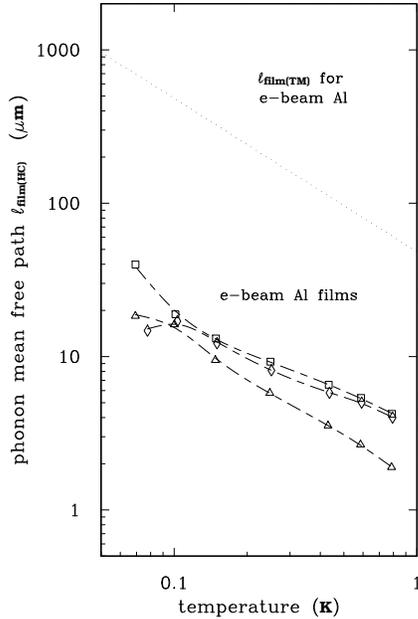}
\caption[a]{Phonon mean free path $\ell_{\rm film(HC)}$ for {\em
e}-beam Al films that are 0.2~$\mu$m (open triangles), 0.4~$\mu$m
(open squares), and 0.6~$\mu$m (open diamonds) thick. The dotted
line is the TM prediction based on internal friction (see Table
III). Dashed lines are guides for the eye.} \label{fig8}
\end{center}
\end{figure}
} earlier\cite{33} (see also Table III).  As for the two previous
examples, the observed phonon scattering far exceeds the
scattering expected on the basis of the TM.

In Ref.\cite{33}, it had been shown that the low temperature
internal friction of an Al film on a Si substrate was very
similar to that of heavily deformed bulk Al.  It was therefore
interesting to compare the phonon mean free  path in the film
with that observed in deformed bulk Al. For that purpose, a
99.999\% pure polycrystalline Al rod (2.5~mm in diameter and
25.7~mm long) was first annealed at 560$^\circ$C and subsequently
stretched by 5\%. Its thermal conductivity, measured by the
standard steady-state technique, is shown in Fig.~\ref{fig9}
along with that of bulk {\em a}-SiO$_2$. The steep rise of the
thermal conductivity of the bulk Al above 0.1~K is caused by the
onset of heat transport by normal state electrons. However, below
that temperature, heat is expected to be carried predominantly by
the lattice, and a temperature dependence similar to that of the
bulk glass ({\em a}-SiO$_2$) is observed. Although the magnitude
is three times smaller, it still falls within the glassy range in
thermal conductivity, see Fig.~1 in ref. \cite{thompson-2001}.
The phonon thermal conductivity of the 0.2~$\mu$m {\em e}-beam Al
film, as calculated from $\ell_{\rm film(HC)}$ in
Fig.~\ref{fig8}, is also shown in Fig.~\ref{fig9}. Above 0.1~K,
the thermal conductivity of the deformed bulk Al and the phonon
thermal conductivity of the {\em e}-beam Al film show the
difference between the heat transport by electrons and phonons,
and by the phonons alone, separated here experimentally for the
first time. Below 0.1~K, the phonon thermal conduc-\vbox{
\begin{figure}[!h]
\begin{center}
\includegraphics[scale=0.5]{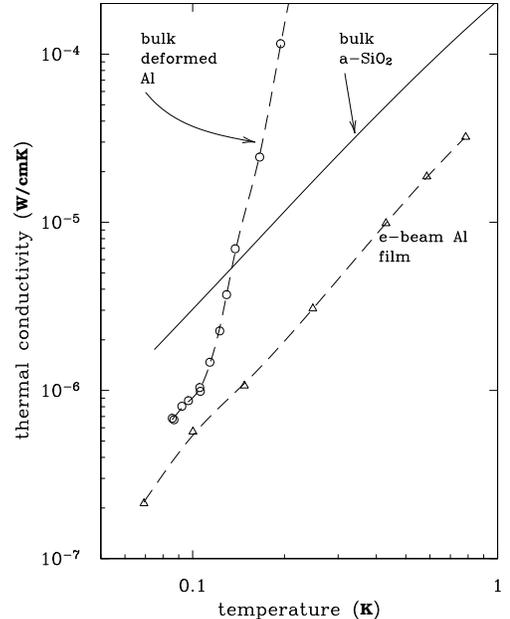}
\caption[a]{The thermal conductivity of 5\% deformed bulk Al. The
thermal conductivity of a bulk {\em a}-SiO$_2$, taken from
Ref.\cite{15}, along with the phonon thermal conductivity of the
0.2~$\mu$m thick {\em e}-beam Al film converted from their phonon
mean free path shown in Fig.~\ref{fig8}, is shown for
comparison.} \label{fig9}
\end{center}
\end{figure}
} tivity of the {\em e}-beam Al film is very close to that of the
deformed bulk sample. This suggests that the defects which
scatter the phonons in the film are very similar to those in the
heavily deformed bulk sample. The same conclusion had been
reached previously in internal friction measurements as stated
above. The defects causing the internal friction had been
tentatively identified as dislocations or dislocation
kinks\cite{33}. It is tempting to suggest that the thermal
phonons in the films are scattered by the same defects. Since we
see in Fig.~\ref{fig8} that $\ell_{\rm film(HC)}$ and $\ell_{\rm
film(TM)}$ of the {\em e}-beam Al films are not connected by the
TM, we can conclude that the same holds for the deformed bulk Al
because of the similarity in the internal friction and phonon
mean free path between the thin films and the bulk samples. Thus,
the non-glasslike phonon scattering phenomena observed in this
work are not limited to thin films alone. In addition, the
defects or the mechanisms causing the resonant scattering of
thermal phonons in heat conduction and those leading to the
relaxational process in internal friction may not even be
related, as shown by the following observation.

The alloy Al 5056 in bulk form has an exceptionally small low
temperature internal friction, even as a sputtered film (Table
III), which has been explained by dislocation pinning\cite{33}.
The $\ell_{\rm film(HC)}$ in this film, however, is still close to
that of all other metal films, see Fig.~\ref{fig10}. It follows
that pinning of dislocations has no influence on the thermal
\vbox{
\begin{figure}[!h]
\begin{center}
\includegraphics[scale=0.5]{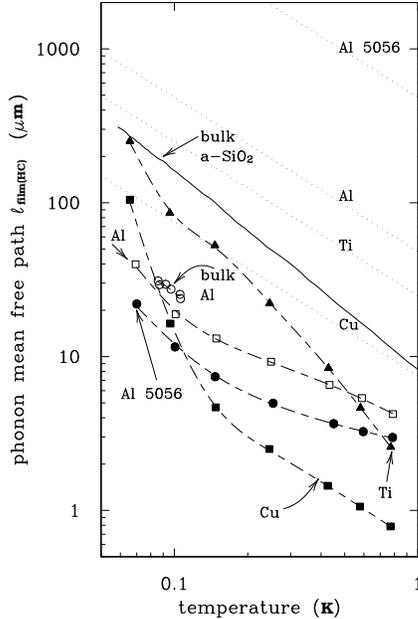}
\caption[a]{Phonon mean free path $\ell_{\rm film(HC)}$ for a
0.4~$\mu$m thick {\em e}-beam Al film: open squares; for a
0.5~$\mu$m thick sputtered alloy Al 5056 film: solid circles; for
a 0.1~$\mu$m thick {\em e}-beam Ti film: solid triangles; and for
a 0.1~$\mu$m thick {\em e}-beam Cu film: solid squares. The
thermal conductivity of for the 5\% deformed bulk Al is converted
to its mean free path: open circles. The labelled dotted lines are
TM predictions of $\ell_{\rm film(TM)}$ based on internal
friction measurements (see Table III). The solid curve is the
phonon mean free path in bulk {\em a}-SiO$_2$ taken from
Ref.\cite{15}. Dashed lines are guides for the eye.} \label{fig10}
\end{center}
\end{figure}
} phonon scattering. We conclude that the mechanisms causing the
internal friction and the thermal phonon scattering are not
understood.

A comparison between $\ell_{\rm film(HC)}$ and $\ell_{\rm
film(TM)}$ is shown in Fig.~\ref{fig10} for four metal films: Al,
alloy Al 5056, Ti, and Cu, with the phonon mean free path of {\em
a}-SiO$_2$ for comparison. The apparent lack of correlation
between $\ell_{\rm film(HC)}$ and $\ell_{\rm film(TM)}$ enable us
to generalize the same conclusion from the Al films to other
metallic films, which is that if glass-like lattice vibrations
exist in them, their effect is masked by the unknown defects.

As observed in internal friction\cite{33}, $\ell_{\rm film(HC)}$
is unaffected by superconductivity ($T_c$ is 0.4~K for Ti, 0.92
for alloy Al 5056\cite{duffy-1990}, and 1.2~K for Al, ). It is
concluded that phonon scattering by electrons is unimportant.
Klemens has derived an expression for the phonon-electron
scattering coefficient $P$ in terms of the electron-phonon
scattering coefficient $E$\cite{83}. Using the value for $E$
measured by Berman and MacDonald for pure copper\cite{84,85}, we
calculate $\ell_{\rm film}$ (of phonons being scattered by
electrons) at 1~K to be $\sim 15$~$\mu$m. This phonon scattering
rate (due to electrons) is more than an order of magnitude less
than the phonon scattering rate observed in Fig.~\ref{fig10} for
the Cu film. Note that the calculation of 15~$\mu$m should not be
taken too seriously as its assumptions\cite{86} of the adiabatic
principle, of a phonon Debye spectrum, and of a free electron gas
may not be adequate at these temperatures for a thin
polycrystalline Cu film with a residual resistivity ratio of
2\cite{33}.  Nevertheless, this estimate agrees with our
observation that electron-phonon interaction is not significant
in our experiment.

\section{Conclusions}

Measurements of the thermal phonon mean free path on films of
amorphous and MBE Si, of polycrystalline and MBE CaF$_2$, of pure
metallic Al, Cu, and Ti, and of the metallic alloy Al 5056 below
1.0~K have revealed, in all cases, similar strong phonon
scattering. Scattering by surface and interface roughness can be
excluded, since nearly the same $\ell_{\rm film(HC)}$ has been
observed in Al films of different thicknesses.  In searching for
the origin of this phonon scattering, we have also measured the
low temperature internal friction of the Si and CaF$_2$ films
(that of the metal films had been measured previously,
Ref.\cite{33}) and also the thermal conductivity of a bulk Al rod
after a 5\% plastic elongation.  In all cases, $\ell_{\rm
film(HC)}$ was found to be much smaller than $\ell_{\rm
film(TM)}$ based on the internal friction and assuming that the
lattice vibrations are glass-like. The discrepancy is
particularly striking for the MBE Si film in which no internal
friction was observed, yet $\ell_{\rm film(HC)}$ was similar to
that found in all other films.  In this case, the phonon
scattering is particularly puzzling since the film is expected to
be structurally more perfect.  In all other films, macroscopic
defects like grain boundaries, voids, cracks, or dislocations may
be the cause for the phonon scattering.  In the deformed bulk Al,
the phonon mean free path was found to be equal to that in thin
Al films.  Since in the bulk sample, individual dislocations or
aggregates thereof are likely phonon scatterers, they may also be
the cause for the scattering in the films. However, dislocation
motion, presumably tunneling, which has been invoked to explain
the internal friction of deformed Al and of Al films (see
Ref.\cite{33,103}) is an unlikely cause for the thermal phonon
scattering since the same $\ell_{\rm film(HC)}$ in Al was also
found in the alloy Al 5056, in which dislocation motion appears
to be suppressed, resulting in a greatly reduced internal
friction.  In conclusion, both internal friction and phonon
scattering have been shown to be sensitive probes for thin film
disorder, including that in MBE Si.  The nature of such disorder
and the mechanisms by which it affects the elastic and thermal
properties are completely unknown. No evidence for the existence
of glass-like lattice vibrations has been detected.

\vskip .20in \begin{center}{\large \bf Acknowledgements}
\end{center} \vskip .10in

We gratefully acknowledge the help of Aaron Judy with the AFM
measurements, Glen Wilk in preparing the MBE Si film at Texas
Instruments (Dallas), and Ken Krebs in fabricating the MBE
CaF$_2$ film at the University of Georgia (Athens) and in
providing very useful information on that film's defects. We
thank Mauro Sardela and David Cahill for XRD analysis at the
University of Illinois (Champaign-Urbana) and Bob Reedy for the
SIMS investigation of the MBE Si film at the National Renewable
Energy Laboratory.  We also thank R.S. Crandall for fruitful
discussions. This work was supported by the National Science
Foundation, Grant No. DMR--9701972, the National Renewable Energy
Laboratory, Grant No. RAD-8-18668, and the Naval Research
Laboratory. Additional support was received from the Cornell
Nanofabrication Facility, NSF Grant No. ECS--9319005, and the
Cornell Center for Materials Research, Award No. DMR-9121564.

\vskip .20in \begin{center} {\large \bf Appendix}
\end{center} \vskip .10in

The technique used in this investigation for the measurement of
the thermal phonon mean free path in thin films has been
described before\cite{91}.  Although the experimental schematic
resembles that of a thermal conductivity measurement, see
Fig.~\ref{fig11}, it should be emphasized that our experiment
leads directly to a thermal phonon mean free path, rather than to
a thermal conductivity $\Lambda$, from which the mean free path
$\ell$ has to be calculated using the gas-kinetic expression
\begin{equation}
\Lambda = \frac{1}{3} C_v \bar\upsilon \ell,
\label{eq:2}
\end{equation}
which requires knowledge of the specific heat $C_v$ of the heat
carrying excitations or phonons traveling with an average
velocity $\bar\upsilon$. In amorphous solids, for example, this
$C_v$ cannot be measured. It can only be calculated from
$\bar\upsilon$ through the use of the Debye model.

The analysis of the heat conduction measurements on the silicon
substrate carrying the film requires a Monte Carlo simulation
which, though straightforward, is nonetheless time-consuming.  By
strictly adhering to the specifics as listed below (including
sample and clamp geometry), one can extract the phonon mean free
path in a film, $\ell_{\rm film}$ (called $\ell_{\rm film(HC)}$
in this paper), from $\ell_{\rm exp}$ (the experimentally
measured phonon mean free path of the film-substrate sample)
without having to repeat any MC simulations, as will be shown in
this Appendix.

Fig.~\ref{fig12} is a plot of the results of MC simulations on a
film-substrate sample with dimensions typically used in this
investigation. For any particular simulation, $\ell_{\rm film}/x$
is an input parameter where $x$ is the film thickness.  The
output parameter is $\ell_{\rm MC}$, a simulated phonon mean free
path of the film-substrate sample.  To determine $\ell_{\rm
film}$, one sets an $\ell_{\rm exp}$ equal to an $\ell_{\rm MC}$
in Fig.~\ref{fig12} to find the \vbox{
\begin{figure}[!h]
\begin{center}
\includegraphics[scale=0.5]{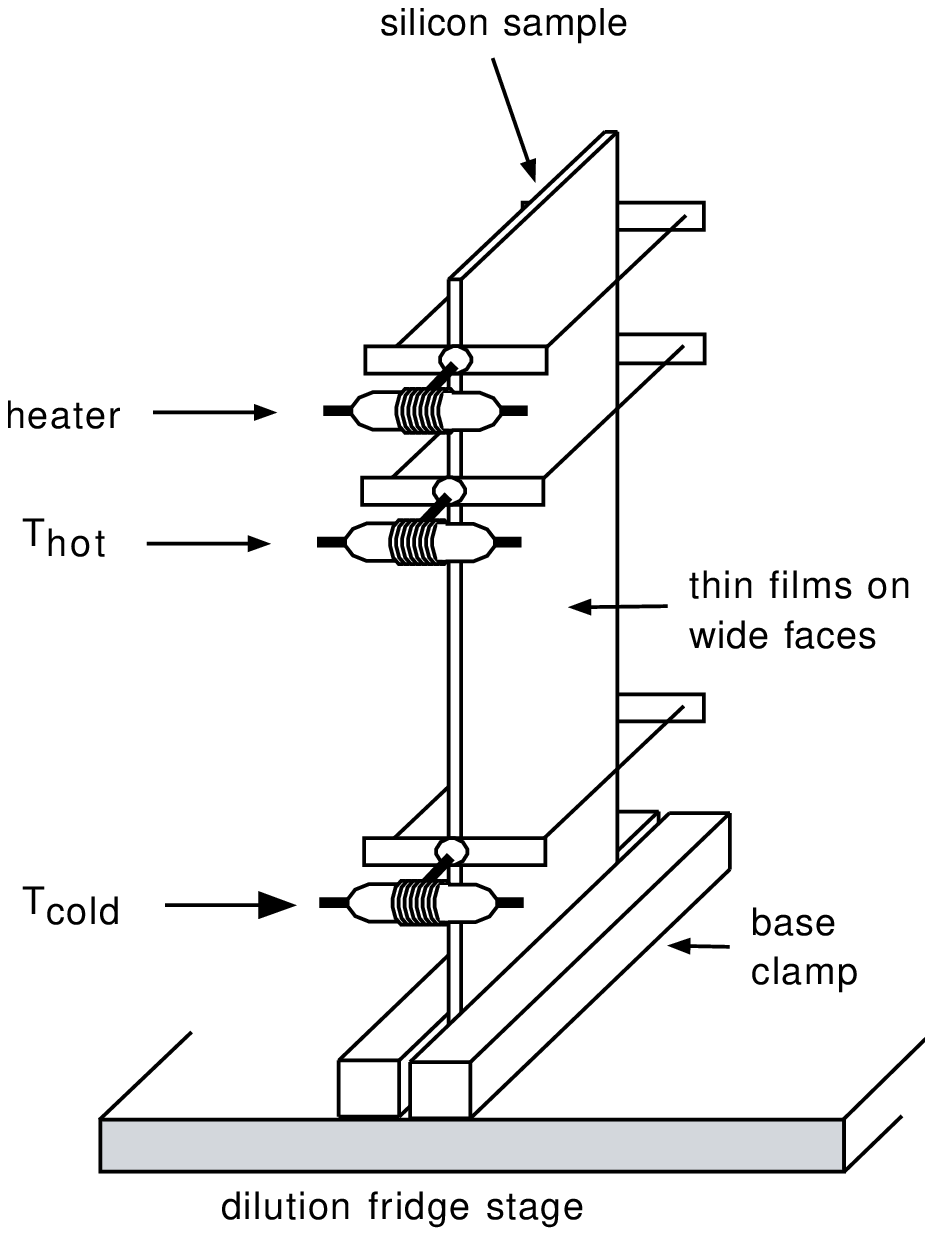}
\includegraphics[scale=0.5]{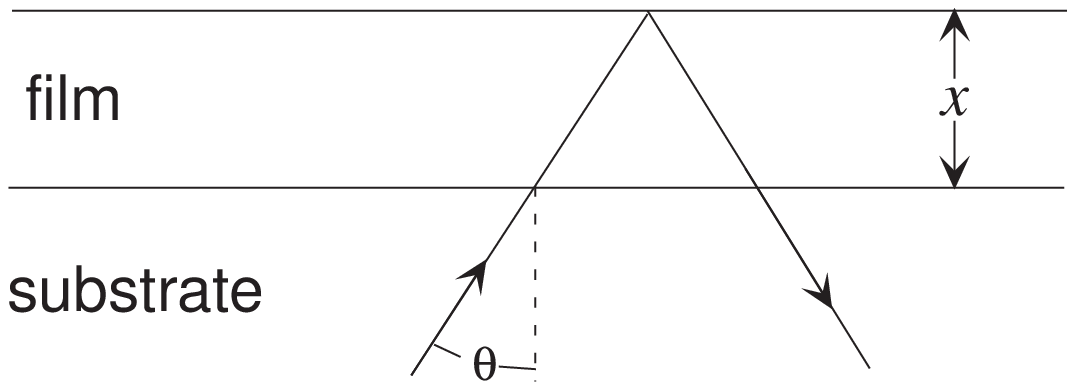}
\caption[a]{Schematic of 1-D heat conduction experiment; the
sample is cleaved from a high purity commercial silicon wafer
(orientation $\langle 111 \rangle$ or $\langle 100 \rangle$) with
both large faces polished; the thin faces are sandblasted as
described in Ref.\cite{91}. Also shown is a ballistic path of a
thermal phonon from the silicon substrate through a thin film as
modeled in the MC simulations.} \label{fig11}
\end{center}
\end{figure}
\begin{figure}[!h]
\begin{center}
\includegraphics[scale=0.5]{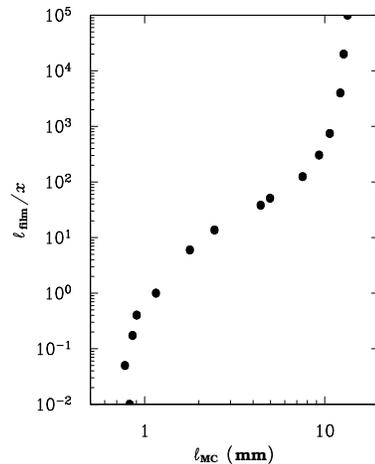}
\caption[a]{Typical plot of $\ell_{\rm film}/x$ versus $\ell_{\rm
MC}$ where $x$ is the film thickness.  Normalizing $\ell_{\rm
film}$ with respect to $x$ allows the same plot to be used for
different film thicknesses as long as the sample and clamp
geometry has not changed (see text for details).  The actual code
of the Monte Carlo programs may be found in
Ref.\cite{92}.}\label{fig12}
\end{center}
\end{figure}
}
corresponding $\ell_{\rm film}/x$; multiplying by $x$ then
yields the desired value, $\ell_{\rm film}$, that directly
corresponds to the measured $\ell_{\rm exp}$.

In order to use Fig.~\ref{fig12} as above, the film-substrate
sample must be mounted as shown in Fig.~\ref{fig11}. Furthermore,
the four thin sides of the high purity silicon substrate must be
completely roughened (by sandblasting, for example) while the two
wide faces must be smooth from Syton polishing, for example), as
explained in Ref.\cite{91}. Any film of thickness $x$ must cover
both wide faces entirely and uniformly. The height of the sample
above the top of the base clamp should be 44.5~mm, the width of
the sample 7~mm, and the thickness of the substrate 0.279~mm. The
height of the heater, cold thermometer, and hot thermometer clamp
above the top of the base clamp should be 41~mm, 4~mm, and 25~mm,
respectively. The two interfaces between the heater clamp and the
sample (areas of contact) should each be of dimensions
$0.279\times 2$~mm$^2$ while the four interfaces between the
thermometer clamps and the sample should each be $0.279\times
1.5$~mm$^2$. The effect of varying these details are discussed in
Ref.\cite{91}.

\bibliography{phonon2kv1}

\end{multicols}

\end{document}